\documentclass[%
twocolumn,
 amsmath,amssymb,
 aps, citeautoscript,
prb,
]{revtex4-1}

\usepackage{graphicx}
\usepackage{dcolumn}
\usepackage{bm}

\begin{document}

\preprint{APS/123-QED}

\title{Large Disparity Between Optical and Fundamental Band Gaps in Layered In$_2$Se$_3$ }

%

\author{Wei Li$^{a}$}
\author{Fernando P. Sabino$^{a}$}
\author{Felipe Crasto de Lima$^{a,b}$}
\author{Tianshi Wang$^{a}$}
\author{Roberto H. Miwa$^{b}$}
\author{Anderson Janotti$^{a}$}
\affiliation{$^{a}$Department of Materials Science and Engineering, University of Delaware, Newark, DE 19716}
\affiliation{$^{b}$Instituto de F\'isica, Universidade Federal de Uberl\^andia, C. P. 593, 38400-902, Uberl\^andia, MG, Brazil}

\date{\today}
\begin{abstract}

In$_2$Se$_3$ is a semiconductor material that can be stabilized in different crystal structures (at least one 3D and several  2D layered structures have been reported) with diverse electrical and optical properties.  This feature has plagued its characterization over the years, with reported band gaps varying in an unacceptable range of 1 eV.  Using first-principles calculations based on density functional theory and the HSE06 hybrid functional, we investigated the structural and electronic properties of four layered phases of In$_2$Se$_3$,  addressing their relative stability and the nature of their fundamental band gaps, i.e., direct {\em versus} indirect. Our results show large disparities between fundamental and optical gaps. 
The absorption coefficients are found to be as high as that in direct-gap III-V semiconductors.  The band alignment with respect to conventional semiconductors indicate a tendency to $n$-type conductivity, explaining recent experimental observations.

\end{abstract}

\maketitle


\section{Introduction}

Chalcogenides form a large family of 2D layered materials with diverse electronic and optical properties, that includes metals\cite{Choi2017}, semiconductors \cite{Yang2016,Zheng2016}, and topological insulators \cite{Zhang2009}. As typical of 2D layered materials, their electronic and optical properties strongly depend  on the number of layers, the layer stacking sequence, and how the atoms are arranged within each layer \cite{Padilha2014}. In$_2$Se$_3$  is a distinguished member of this family of compounds.  It has been investigated for many technological applications, including solar cells \cite{Peng2007}, photodetectors \cite{Balakrishnan2016,Zhai2010,Jacobs-Gedrim2014a,Island2015} and phase-change memory devices \cite{Lee2005,Yu2007}. Extraordinary photoresponse in 2D In$_2$Se$_3$ nanosheets has been observed \cite{Jacobs-Gedrim2014a,Zheng2015}, with key figures of merit exceeding those of graphene and other 2D materials based photodetectors; the reported photoconductive response extends into ultraviolet, visible, and near-infrared spectral regions. In$_2$Se$_3$-based phase-change memories have been demonstrated, exploring transitions between different polytypes with diverse electrical properties \cite{Lee2005,Yu2007}.  More recently, ferroelectric ordering in 2D In$_2$Se$_3$ has also been predicted \cite{Ding2017}, creating prospects of room-temperature ferroelectricity with reversible spontaneous electric polarization in both out-of-plane and in-plane orientations.  All these properties and potential applications are affected by or depend on the polymorphism of In$_2$Se$_3$.  The ease of stabilizing In$_2$Se$_3$ in different crystal structures  with diverse electronic and optical properties can be detrimental to photodetectors, yet it may be desirable for phase-memory devices where the involved structures must display disparate electrical properties.  

\begin{figure*}[ht!]
\begin{center}
\includegraphics[width=6.0 in]{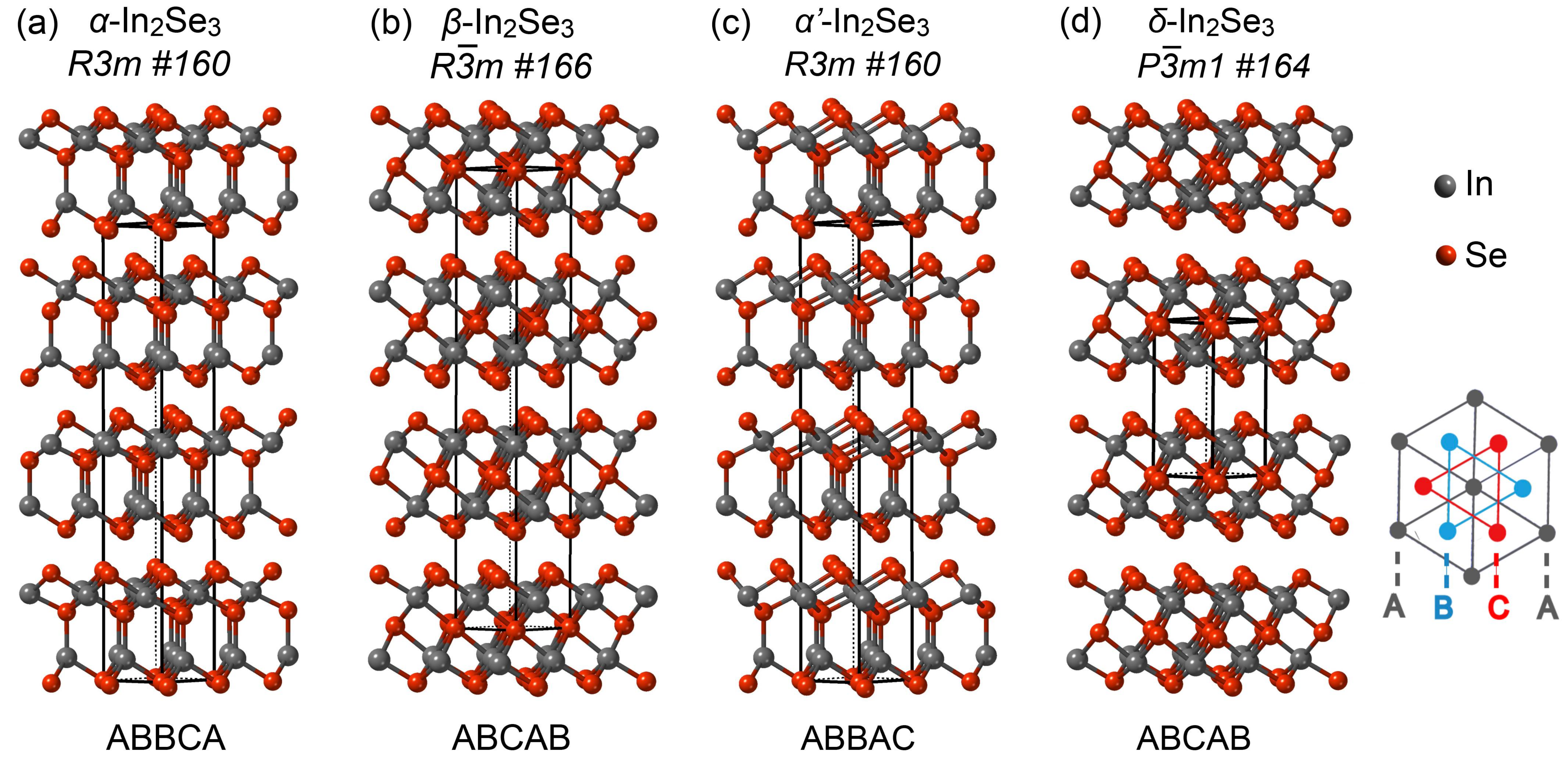}
\end{center}
\caption{Ball \& stick model of the crystal structures of layered In$_2$Se$_3$ considered in the present work: (a) $\alpha$-In$_2$Se$_3$, (b) $\beta$-In$_2$Se$_3$, (c) $\alpha'$-In$_2$Se$_3$, 
and (d) $\delta$-In$_2$Se$_3$.  The space groups and the stacking within each quintuple layer are indicated. For $\alpha$, $\beta$ and $\alpha'$-In$_2$Se$_3$, the hexagonal unit cells containing three formula units are displayed.  For $\delta$-In$_2$Se$_3$, the hexagonal cell shown is the primitive cell and contains two formula units.
}
\label{fig1}
\end{figure*}

Although In$_2$Se$_3$ has been studied for many years, the reports on crystal structure are rather confusing and even contradictory in many cases \cite{Osamura1966,Popovic1979,Ye1998}, with remarkable disagreements on atomic positions within the layers and layer stacking sequence. At least four phases have been reported ($\alpha$, $\beta$, $\gamma$, and $\delta$), with one of them being a 3D phase (labeled $\gamma$) and the others identified as layered phases. The layered structures are composed of five atomic layer Se-In-Se-In-Se sets, with strong covalent bonds within each quintuple layer and van der Waals interactions connecting neighboring quintuple layers.  Among the layered phases of In$_2$Se$_3$,  $\alpha$ and $\beta$, shown in Fig.~\ref{fig1}(a) and (b), are the most prominent, with a reported $\alpha$$\rightarrow$$\beta$ transition temperature of 473 K \cite{Ei-Shair1992}.


The reported values for the band gap of layered In$_2$Se$_3$, either from optical absorption spectra \cite{Marsillac2011,Ho2013,G2009,Bodnar2016}  or calculated using first-principles methods based on the density functional theory \cite{Debbichi2015,Ji2013}, vary from 0.55 eV to 1.5 eV, and the nature of the gap, i.e., direct or indirect, has often been overlooked. Here we perform hybrid functional calculations for the electronic and optical properties of four layered structures of In$_2$Se$_3$ (including the $\alpha$ and $\beta$ structures shown in Fig.~\ref{fig1}), paying special attention to the disparity between optical and fundamental band gaps.  We compare the stability of the different phases through their formation enthalpies, calculate the real and imaginary parts of the dielectric functions, and determine optical absorption coefficients.    We find that the layered structures all have indirect band gaps, with the lowest band gap of 0.17 eV and the highest of 1.35 eV.  The calculated optical transition matrix elements reveal that the optical gap is significantly different from the fundamental band gap for two of the structures, and that the onset of optical absorption all occur at energies higher than 1 eV.  We also compute the band alignment between the different phases, and find that
the position of the conduction-band minimum (CBM) is relatively low with respect to the vacuum level, indicating a tendency for $n$-type conductivity for all the layered In$_2$Se$_3$ structures.

\section{Computational approach}

The calculations are based on the density functional theory \cite{Kohn1964,Kohn1965} and the screened hybrid functional of Heyd-Scuseria-Ernzerhof (HSE06) \cite{Heyd2003,Heyd2006} as implemented in the VASP code \cite{Kresse1993a,Kresse1993b}. The interactions between the valence electrons and the ions are described using projector augmented wave (PAW) potentials \cite{Blochl1994,Kresse1999}.   To improve the description of the weak interaction between the quintuple layers of  In$_2$Se$_3$, we adopted a van der Waals (vdW) correction according to DFT-D2 method of Grimme \cite{JCC:JCC20495}. The structures were optimized using a cutoff of 320 eV for the plane wave basis set, until forces on the atoms were lower than 0.005 eV/\AA. The Brillouin zone was sampled using a $\Gamma$-centered 6$\times$6$\times$6  mesh of $k$-points for the primitive cells.

\begin{figure}[ht]
\begin{center}
\includegraphics[width=3.2 in]{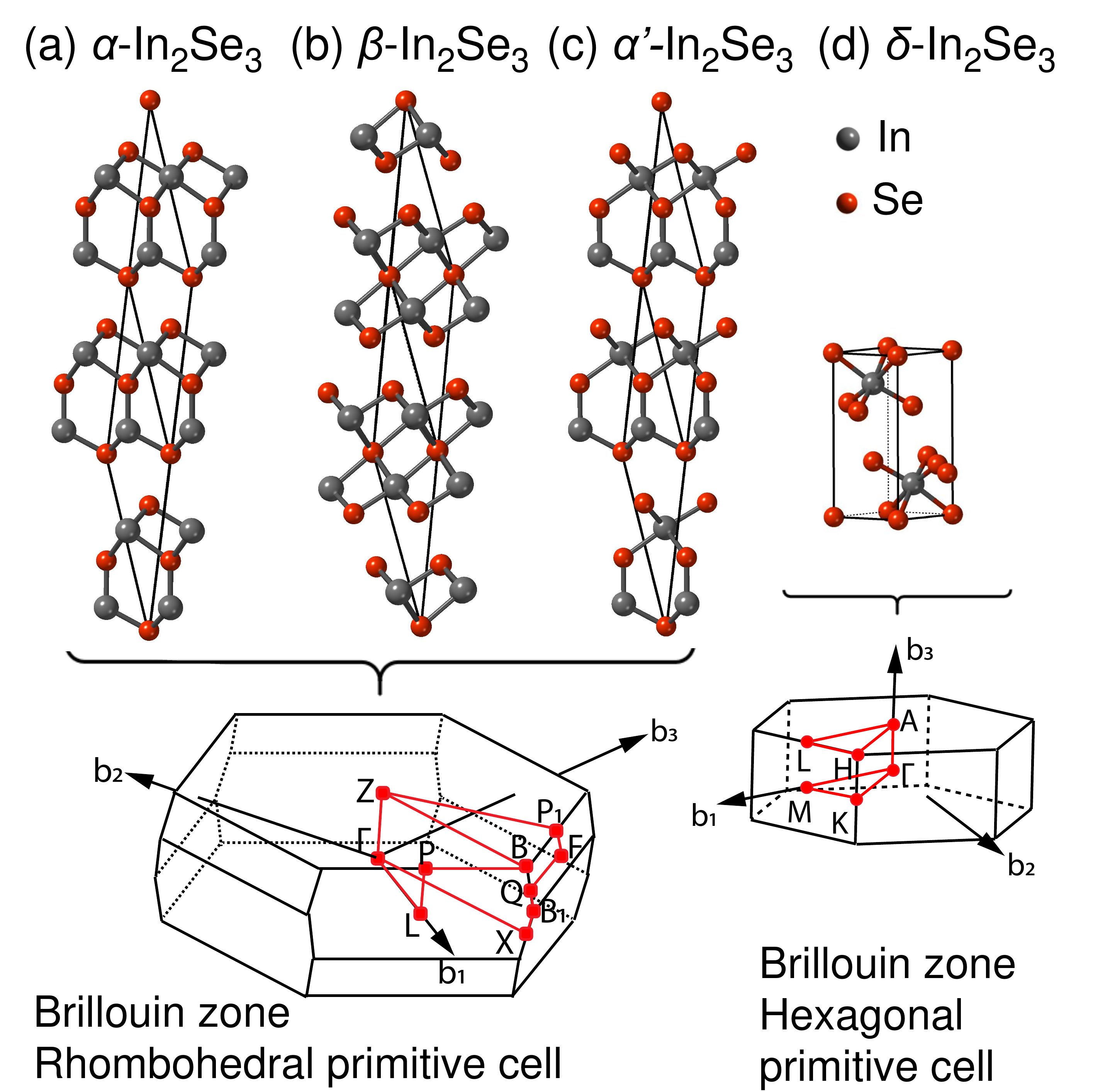}
\end{center}
\caption{Rhombohedral primitive cells for (a) $\alpha$, (b) $\beta$, and (c) $\alpha'$-In$_2$Se$_3$, and the hexagonal primitive cell for (d) $\delta$-In$_2$Se$_3$. The Brillouin zones are shown at the bottom, along the $k$-paths used to plot the band structures.
}
\label{fig2}
\end{figure}

The structures of In$_2$Se$_3$ in Fig.~\ref{fig1}(a)-(c) can be described by rhombohedral primitive cells containing one formula unit, while the structure in Fig.~\ref{fig1}(d) is described by a hexagonal primitive cell with two formula units. These primitive cells are shown in Fig.~\ref{fig2}.  For the rhombohedral primitive cells we chose the following lattice vectors\cite{Furthmuller2016}:
\begin{align}
\vec{a}_1=(b',a',a'); \nonumber  \\
\vec{a}_2=(a',b',a');  \nonumber \\
\vec{a}_3=(a',a',b')\text{,}
\label{LV}
\end{align}
where the three vectors have the same length $a=\sqrt{2a'^2+b'^2}$, and form an angle $\theta$ defined by $\theta$=$\arccos {[(a'^2+2a'b')/a^2]}$.  In practice, 
the lattice parameters of the layered structures of In$_2$Se$_3$ are often reported using conventional hexagonal unit cells.  
Our choice of lattice vectors for the rhombohedral primitive cells makes it easy to express the lattice parameters of the hexagonal unit cells,
$a_{hex}$ and $c_{hex}$ in terms of $a$ and $\theta$ above.  The lattice vectors of the hexagonal unit cells, containing three formula units, 
are given by:
\begin{align}
a_{hex} &=a\sqrt{2(1-\cos \theta)} \nonumber \\
c_{hex} &=a\sqrt{3(1+2 \cos \theta)}\text{,}
\end{align} 

The dielectric function along the in-plane $a_{hex}$ direction ($\varepsilon_{hex,\parallel}$) and out-of-plane $c_{hex}$ direction ($\varepsilon_{hex,\bot}$) are written as:
\begin{align}
\varepsilon_{hex,\parallel}     & =\varepsilon_{diag}-\varepsilon_{nondiag} \nonumber \\
\varepsilon_{hex,\bot} & =\varepsilon_{diag}+2\varepsilon_{nondiag}, 
\label{EPS}
\end{align}
where $\varepsilon_{nondiag}$ and $\varepsilon_{diag}$ are the diagonal and nondiagonal elements of the dielectric tensor obtained using the rhombohedral primitive cells with the lattice vectors given by Eq.~\ref{LV} \cite{Furthmuller2016}.
The electronic band structures and dielectric functions were calculated using the HSE06 hybrid functional. Calculations using the GW method \cite{Luoie1996,Kresse2006} give band gaps that are systematically higher by only 0.1 eV.  Contributions from excitons and phonon-assisted optical transitions to the absorption coefficient are expected to be relatively small and were not included in the present work. 

\section{Results and discussion}

\subsection{Structure and stability of layered  In$_2$Se$_3$}

\begin{table}[ht]
\caption{Calculated lattice parameters and formation enthalpies $\Delta$H$_f$ for layered phases of In$_2$Se$_3$ using HSE06 and HSE06+vdW.
Previous experimental and theoretical results are also listed for comparison.}
\label{tbl:table1}
\begin{tabular}{llll}
\hline
&$a_{hex}$ (\AA) & $c_{hex}$ (\AA)&$\Delta$H$_f$(eV)\\
\hline
\vspace{1mm}
$\alpha$-In$_2$Se$_3$&\\
HSE06& 4.066 & 30.368 & -3.982 \\
HSE06+vdW& 3.973 & 28.752 & -3.109 \\
Exp.  &  4.00 & 28.80 \cite{Ye1998} & -2.858 \cite{Chatillon1993}  \\
& 4.05 & 28.77 \cite{Osamura1966} &\\
Previous calc. & 3.93 & 27.9 \cite{Debbichi2015} &\\
\hline
$\beta$-In$_2$Se$_3$&\\
HSE06 & 3.978 & 29.890 & -3.737\\
HSE+vdW & 3.904 & 27.671 & -3.071\\
Exp.  & 4.025 & 28.762 \cite{Lutz1988} &\\
& 4.05 & 29.41\cite{Osamura1966} &\\
Previous calc.& 4.00 & 29.04 \cite{Debbichi2015}  &\\
\hline
$\delta$-In$_2$Se$_3$&\\
HSE06 & 3.978 & 10.195 &-3.732 \\
HSE06+vdW & 3.902 & 9.322 &-3.042 \\
Exp.  & 4.01 & 9.64 \cite{Popovic1979} & \\
\hline
$\alpha'$-In$_2$Se$_3$&\\
HSE06& 4.003 & 30.279 & -3.975 \\
HSE06+vdW& 3.975 & 28.785 & -3.103 \\
\hline
\end{tabular} 
\end{table}

The crystal structures  of layered In$_2$Se$_3$ are composed of sets of quintuple layers, Se-In-Se-In-Se, with each atomic layer containing only one elemental species arranged in a triangular lattice. Within the quintuple layers, the atoms form strong covalent/ionic bonds, while the interactions between neighboring quintuple layers are weak and of the van der Waals type. 
The crystal structures in Fig.\ref{fig1} differ in the stacking within the quintuple layer and inter quintuple layers.  The most studied phases of layered In$_2$Se$_3$ are the $\alpha$ and $\beta$ shown in Fig.~\ref{fig1}(a) and (b).  In the $\alpha$-In$_2$Se$_3$ structure, space group $R3m$, the Se-In-Se-In-Se atomic layers are stacked in the ABBCA  sequence, where one of the In is fourfold coordinated in a tetrahedral environment, and the other is sixfold coordinated in an octahedral environment.  In the $\beta$-In$_2$Se$_3$, space group $R\overline{3}m$, both In atoms are sixfold coordinated in octahedral environments.  In a variant of the $\alpha$-In$_2$Se$_3$ structure, here labeled $\alpha'$, space group $R3m$, the Se-In-Se-In-Se atomic layers are stacked in the ABBAC sequence, where one of the In is fourfold coordinated, and the other is sixfold coordinated, as shown in Fig.~\ref{fig1}(c).  The $\delta$-In$_2$Se$_3$ structure, space group $P\overline{3}m_1$, is a variant of the $\beta$-In$_2$Se$_3$, differing only in the stacking of the quintuple layers. While in $\beta$-In$_2$Se$_3$, each period along the out-of-plane direction ($c_{hex}$ axis) contains three quintuple layers [Fig.~\ref{fig1}(b)], in $\delta$-In$_2$Se$_3$, each period along $c_{hex}$ axis contains only one quintuple layer, as shown in  Fig.~\ref{fig2}(d). 

The calculated lattice parameters of $\alpha$, $\beta$, $\alpha'$, and $\delta$-In$_2$Se$_3$, using both HSE06 and HSE06 with van der Waals correction (HSE06+vdW)  are listed in Table~\ref{tbl:table1}. HSE06 leads to a good agreement between theoretical and experimental results for in-plane lattice parameters $a_{hex}$, however the error in the out-of-plane lattice parameter $c_{hex}$ exceeds $8.8\%$ compared to the experimental value for the $\alpha$ structure. HSE06+vdW improves the description of $c_{hex}$, reducing the error to less than  $3.0\%$.

The calculated formation enthalpies $\Delta H_f$ are also listed in Table~\ref{tbl:table1}.  $\Delta H_f$ is defined as:
\begin{equation}
\Delta H_f=E_{tot}({\rm In}_2{\rm Se}_3)-2E_{tot}({\rm In})-3E_{tot}({\rm Se})\text{,}
\label{enthalpy}
\end{equation}
where $E_{tot}({\rm In}_2{\rm Se}_3$) are the total energies per formula unit of In$_2$Se$_3$ in the different crystal structures, $E_{tot}({\rm In})$ and $E_{tot}({\rm Se})$ are the total energies  per atom of In and Se bulk phases.  Using HSE06+vdW, we find that $\alpha$-In$_2$Se$_3$ has the lowest formation enthalpy (-3.109 eV), followed by $\alpha'$, $\beta$, and $\delta$-In$_2$Se$_3$. The calculated formation enthalpy of  $\alpha$-In$_2$Se$_3$ is in good agreement with available experimental data \cite{Chatillon1993}. 
We note that vdW corrections systematically increase formation enthalpies by  about 0.8 eV for the different phases of layered In$_2$Se$_3$.  

\subsection{Electronic structure of layered  In$_2$Se$_3$}

\begin{figure*}[ht!]
\includegraphics[width=5.8 in]{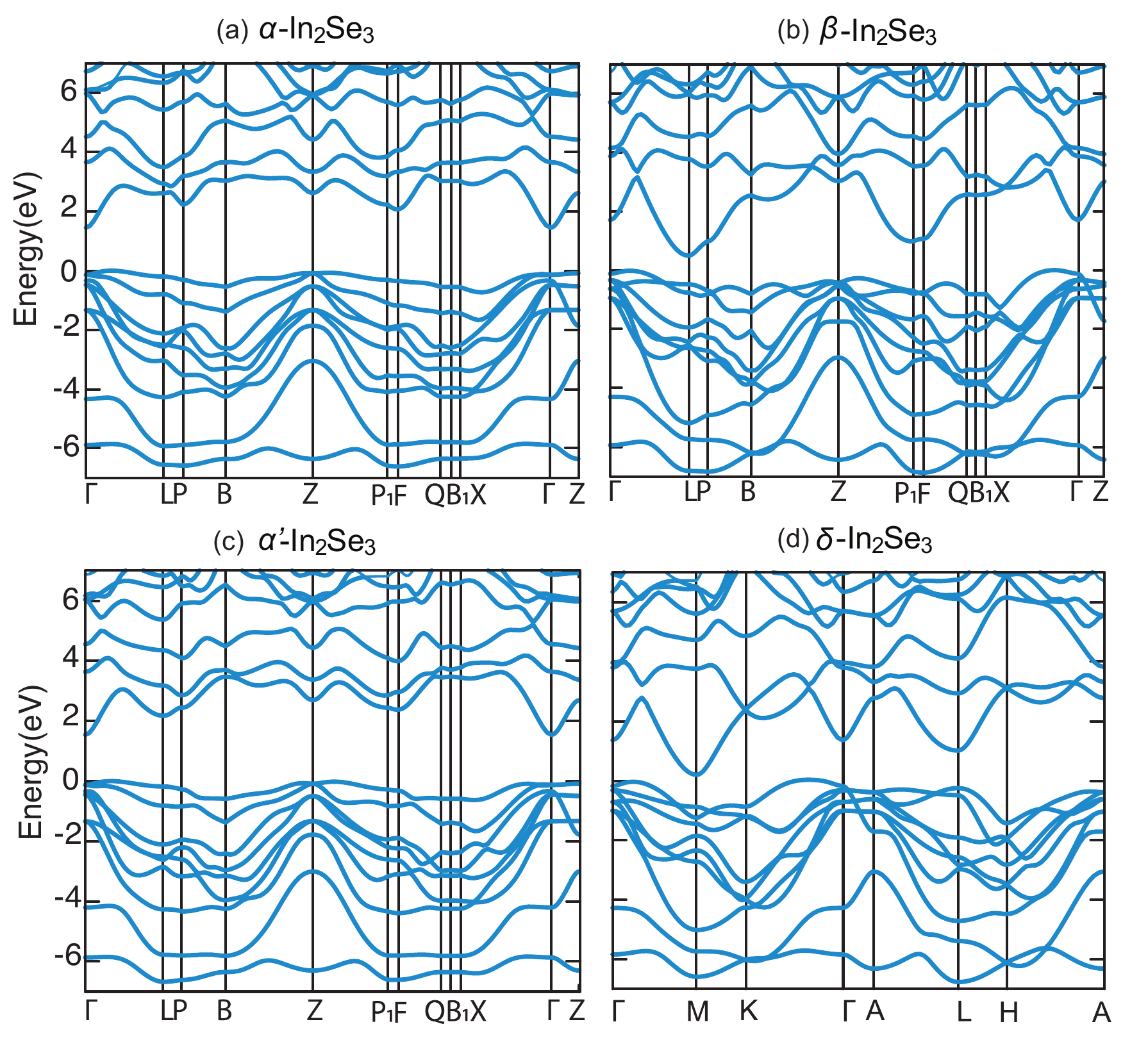}
\caption{Electronic band structure of $\alpha$, $\beta$, $\alpha'$ and $\delta$-In$_2$Se$_3$ using the primitive cells. The valence-band maximum (VBM) is set to 0 eV.} 
\label{fig3}
\end{figure*}

The calculated band structures of $\alpha$, $\beta$, $\alpha'$, and $\delta$-In$_2$Se$_3$, using the primitive cells, are shown in Fig.~\ref{fig3}, including all the high symmetry $k$ points in the irreducible part of the Brillouin zone. The four layered phases of In$_2$Se$_3$ display fundamental indirect band gaps, with highly dispersive conduction bands (small effective electron masses) derived from In $s$ orbitals, and much less dispersive valence bands derived mostly from Se $p$ orbitals. 

For $\alpha$-In$_2$Se$_3$, the valence-band maximum (VBM) occurs along the $\Gamma$-L direction (about one third of the $\Gamma$-L distance from $\Gamma$), while the conduction-band minimum (CBM) is located at $\Gamma$, with an indirect band gap of 1.34 eV.  The direct gap at $\Gamma$ is 1.46 eV. For $\beta$-In$_2$Se$_3$, the VBM occurs along the $\Gamma$-X direction (about one third of the distance from $\Gamma$), while the CBM is located at the L point, with an indirect band gap of 0.49 eV.  The conduction-band edge at $\Gamma$ is 1.21 eV higher than at L, and the valence-band edge at $\Gamma$ is 0.12 eV lower than the VBM along $\Gamma$-X, so that the direct gap at $\Gamma$ is 1.82 eV. The $\alpha'$ phase has a very similar band structure as the $\alpha$-In$_2$Se$_3$, with a band gap of 1.35 eV, i.e., only 0.01 eV higher than that of $\alpha$-In$_2$Se$_3$. The $\delta$-In$_2$Se$_3$ has a small band gap of only 0.17 eV, with the VBM along the $\Gamma$-K direction and the CBM at the M point in the hexagonal Brillouin zone [Fig.~\ref{fig2}(d)]. The direct gap at $\Gamma$ is 1.55 eV.  

Previous results of first-principles calculations for the band gap of In$_2$Se$_3$ vary in a wide range\cite{Debbichi2015,Ji2013}. The few reported band structures for bulk  In$_2$Se$_3$ are nevertheless incomplete for the following reasons. First, the calculations were performed using the hexagonal unit cells, instead of the primitive cells. This may prevent a proper analysis of the direct {\em versus} indirect nature of the band gap since the Brillouin zone of the hexagonal unit cell is folded into that of the rhombohedral primitive cell. A direct gap in the Brillouin zone of the hexagonal cell may well involve distinct $k$ points in the Brillouin zone of the primitive rhombohedral cell. Second, and more worrisome, the calculations for the hexagonal unit cells do not include all high-symmetry $k$ points in the irreducible part of the Brillouin zone; they only include $k$ paths in the in-plane direction passing through the $\Gamma$ point.
In fact, our calculations for the band structure of $\beta$-In$_2$Se$_3$ using the hexagonal unit cell show that while the VBM occurs at the $\Gamma$-K direction, the CBM occurs at the L point, i.e., not located  in $k$ paths in the in-plane direction passing through the $\Gamma$ point [see the Brillouin zone in Fig.~\ref{fig2}(d) for reference].

Based on full-potential linearized augmented plane-wave and local orbitals (FPLAPW+lo) basis method and the modified Becke Johnson (mBJ) meta-GGA, an indirect band gap of 0.55 eV and a direct band gap at $\Gamma$ of 1.5 eV were reported for $\beta$-In$_2$Se$_3$\cite{Ji2013}. The authors argued that the calculated band gap was underestimated due to DFT band gap problem. However, the Becke Johnson (mBJ) meta-GGA approximation was designed to overcome this problem and to give band gaps in close agreement with experimental values.  DFT-GGA calculations for $\alpha$ and $\beta$-In$_2$Se$_3$ resulted in indirect band gaps of 0.49 eV and 0.21 eV, while using the GW method, gaps of 1.25 eV and 0.7 eV were obtained\cite{Debbichi2015}.  However, the authors calculated the band structures using the hexagonal unit cells and only considered in-plane $k$-paths passing through the $\Gamma$ point.

We also calculated the optical transition matrix elements between valence- and conduction-band states for the four layered structures of In$_2$Se$_3$.  
For  the $\alpha$ and $\alpha$' structures, we find the transition at $\Gamma$ to be allowed and slightly higher in energy than the fundamental indirect band gap. For $\beta$-In$_2$Se$_3$, we find that the lowest energy direct transition at L is not allowed and that the optical gap originates from the second valence band to the conduction band at L. 
 For $\delta$-In$_2$Se$_3$, the optical gap is associated with the transition from the first valence band and conduction band at M. These results show a disparity between the fundamental and the optical gaps, which are larger than 1 eV in the case of $\beta$ and $\delta$-In$_2$Se$_3$.

\subsection{Dielectric functions and absorption coefficients of layered  In$_2$Se$_3$}

\begin{figure*}[ht]
\begin{center}
\includegraphics[width=6.4 in]{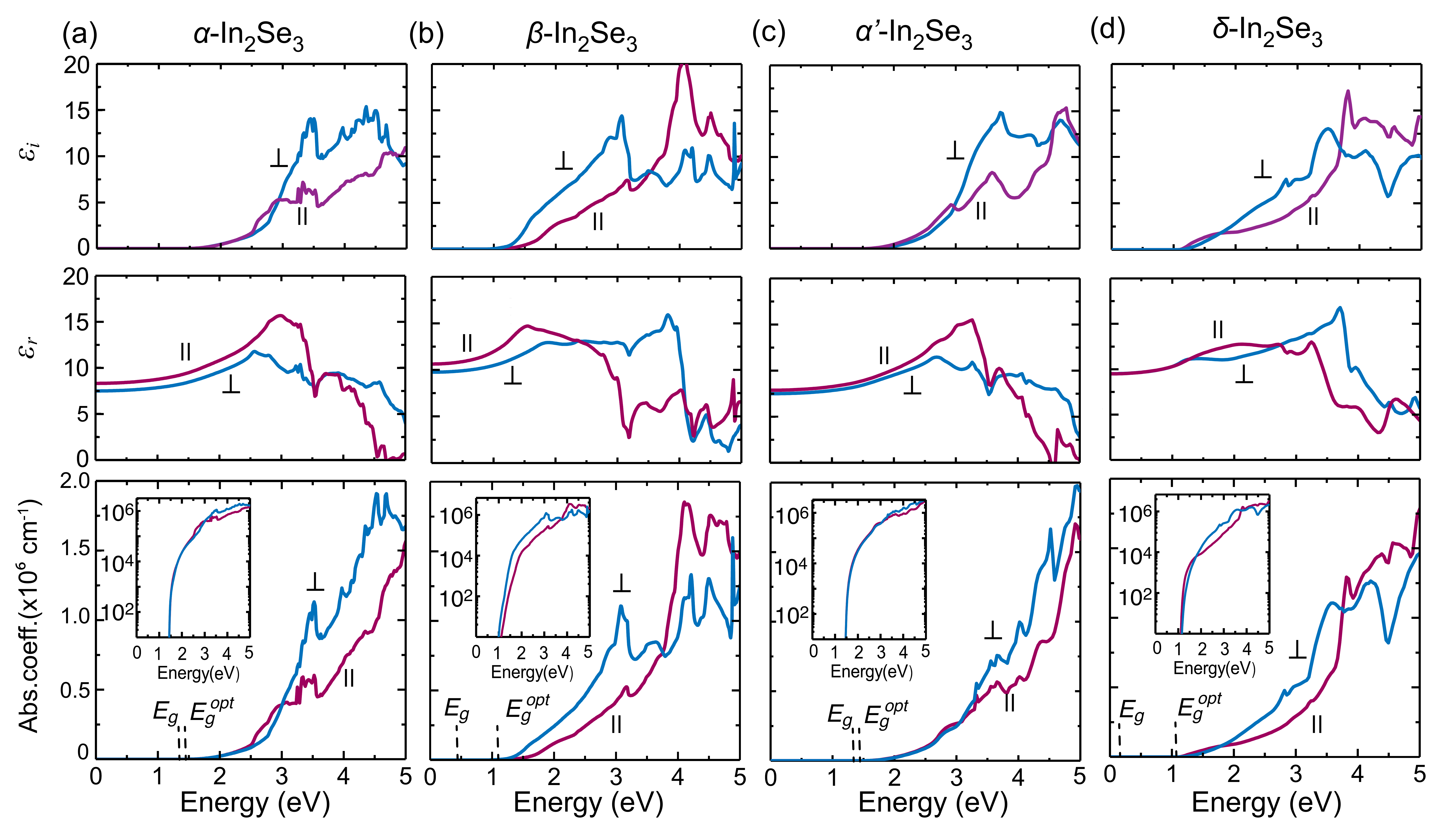}
\end{center}
\caption{Imaginary and real parts of frequency dependent dielectric function, $\varepsilon_i$ and  $\varepsilon_r$, for the layered phases of In$_2$Se$_3$. The calculated absorption coefficients are shown at the lower panels.  The insets use logarithmic scale. The blue lines correspond to light polarization in the out-of-plane direction ($\bot$, along the $c_{hex}$ axis), the purple lines correspond to light polarization in the in-plane direction ($\parallel$, along the $a_{hex}$ axis). of the hexagonal cells of In$_2$Se$_3$.
}
\label{fig4}
\end{figure*}

The optical properties of layered In$_2$Se$_3$ are discussed based on the real and imaginary parts of the dielectric matrix, absorption coefficient and the optical transition matrix elements. 
To determine the optical band gap from the dielectric functions and the derived absorption coefficient we employed the tetrahedral method for the integration over the Brillouin zone, with a small Lorentzian broadening parameter of 0.001 eV.  The calculations for the dielectric function were carried out for the rhombohedral primitive cells for $\alpha$, $\beta$, and $\delta$-In$_2$Se$_3$, and then converted to the hexagonal directions according to Eq.~\ref{EPS}.

The real and imaginary parts of the dielectric function are shown in Fig.~\ref{fig4}. Due to the hexagonal layered structure, we expect the dielectric function to be anisotropic, with nonzero components only in the out-of-plane ($\bot$) and in the in-plane ($\parallel$) directions.

For $\alpha$, $\beta$, and $\alpha'$-In$_2$Se$_3$, the real part of the dielectric tensor at zero energy (or frequency), 
is higher in the in-plane than in the out-of-plane direction, i.e., $\varepsilon^{\infty}_{\parallel}  > \varepsilon^{\infty}_{\bot}$. This is expected since the electronic screening is stronger in the in-plane directions than in the out-of-plane direction due to the layered nature of the crystal structure. For the $\delta$ phase, we find $\varepsilon^{\infty}_{\bot} \approx \varepsilon^{\infty}_{{\parallel}}$, likely due to the alignment of the Se atoms connecting two neighboring quintuple layers, that favors the overlap of Se $p$ orbitals across the quintuple layers, and the smaller distance between the quintuple layers.


The absorption coefficients were determined from the real and imaginary parts of the dielectric matrix and are shown at the bottom panel in Fig.~\ref{fig4}. For all the structures investigated, the amplitude of absorption coefficients are rather large for above  band-gap excitations, i. e., exceeding 10$^6$ cm$^{-1}$, which are of the same order of magnitude as that in direct gap III-V semiconductors \cite{Stillman1984}. The optical absorption coefficients start increasing only at energies higher than 1 eV.  Note that the optical band gap of $\alpha$ and $\alpha'$-In$_2$Se$_3$ are similar to the fundamental band gap, because the small difference between the VBM and the of valence band edge at $\Gamma$-point. In contrast, the optical band gap of 1.27 eV for $\beta$-In$_2$Se$_3$  and  1.11 eV for $\delta$-In$_2$Se$_3$ are much higher than the fundamental gap of 0.49 eV and 0.17 eV, respectively. 

Optical absorption measurements of $\alpha$ and $\beta$-In$_2$Se$_3$ single crystals,
where the $\beta$ form was obtained by heating $\alpha$-In$_2$Se$_3$ crystals above 473 K in a furnace, 
revealed optical band gaps of 1.356 eV and 1.308 eV respectively\cite{Julien1990}. Optical transmission measurements in $\alpha$-In$_2$Se$_3$ films, deposited using thermal evaporation on glass substrates, indicated an indirect fundamental band gap slightly below the optical band gap of 1.37 eV \cite{El-Shair1992}. Layered In$_2$Se$_3$ samples grown by vapor phase technique\cite{Ye1998} have shown onset in the absorption spectra at 1.26 eV, with an estimated fundamental indirect gap of about 1.1 eV.  The authors proposed a structure composed of quintuple layers with an unlikey truncated wurtzite crystal arrangement within each quintuple layer, where one Se atoms at the boundary is one-fold coordinated. We believe these results actually refer to $\alpha$-In$_2$Se$_3$ based on the reported lattice parameters. Films of $\alpha$-In$_2$Se$_3$ fabricated by ion-beam sputtering at 312 K from single crystals showed a band gap of 1.58 eV, determined from transmittance spectra \cite{Bodnar2016}. More recently, a band gap of 1.46 eV for $\alpha$-In$_2$Se$_3$ and 1.38 eV for $\beta$-In$_2$Se$_3$ were determined using photocurrent spectroscopy\cite{Wang2017}. The authors noted that the $\beta$-In$_2$Se$_3$ films ($\sim$86 nm) were highly electrically conductive as a metal. All these results indicate optical band gaps above 1 eV, in agreement with our calculations.

Our calculations also offer some insights on the electronic structure of (In$_x$Bi$_{1-x}$)$_2$Se$_3$ even though we did not carry out explict calculations for this system.
Recent experiments have proposed that adding In to Bi$_2$Se$_3$ increases the band gap and reduces the Fermi level which is resonant in the conduction band of Bi$_2$Se$_3$, leading to a transition from a doped topological insulator to a trivial insulator with band gap over 1 eV \cite{Liu2013}. 
However, it is unclear if the conduction-band edge in the (In$_x$Bi$_{1-x}$)$_2$Se$_3$ alloy is pushed up or if the ARPES measurements do not capture the $k$ range where the band edges occur \cite{Brahlek2012,Wu2013,Liu2013}.  From our results, it is unlikely that alloying $\beta$-In$_2$Se$_3$ (band gap 0f 0.49 eV) with Bi$_2$Se$_3$ (0.22 eV) \cite{Martinez2017} would lead to fundamental band gaps larger than 1 eV as previously proposed \cite{Liu2013}. 

\subsection{Band alignments}

\begin{figure}[ht!]
\begin{center}
\includegraphics[width=3.4 in]{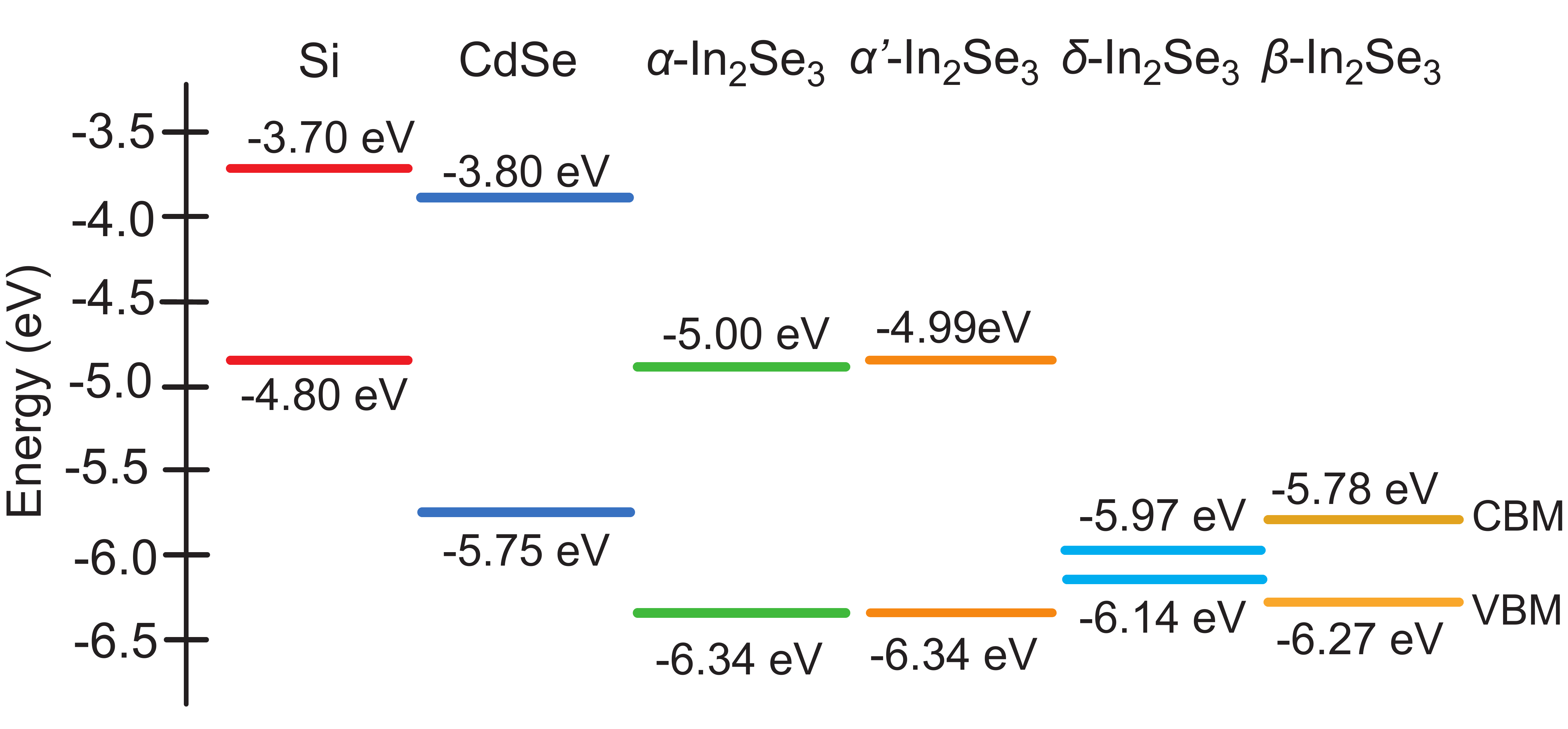}
\end{center}
\caption{Band alignment between $\alpha$, $\beta$, $\alpha'$ and $\delta$-In$_2$Se$_3$ and band edge positions with respect to vacuum level. The position of the band edges of CdSe with respect to vacuum level, extracted from Ref.\cite{VandeWalle2003a} is also shown for comparison. } 
\label{fig5}
\end{figure}

Finally, we calculated the band alignment between the different structures of layered In$_2$Se$_3$.  To be able to directly compare the VBM and CBM of the different phases, we fixed the volume per formula unit in each calculation to be that in $\beta$-In$_2$Se$_3$, so that the average electrostatic potential, which is used as reference in the HSE06 calculations, is the same for the different structures.  In order to align with the vacuum level, we used a slab of $\beta$-In$_2$Se$_3$ with 9 quintuple layers and determined the averaged electrostatic potential of the middle quintuple layer with respect to the potential in the vacuum region of the slab. The results are shown in Fig.~\ref{fig5} and are compared to the values of the VBM and CBM for CdSe and Si from the literature \cite{VandeWalle2003a}.  

From Fig.~\ref{fig5}, we note that the CBM of layered In$_2$Se$_3$ are significantly lower than that of CdSe or Si, and lower than the standard hydrogen electrode potential (i.e., $\sim$-4.5 eV below the vacuum level \cite{VandeWalle2003a}), implying that  all the layered phases of In$_2$Se$_3$ will have a tendency for $n$-type conductivity.  This is specially the case of $\beta$-In$_2$Se$_3$  and $\delta$-In$_2$Se$_3$. This explains the recent measurements of photocurrent spectroscopy \cite{Wang2017} where $\beta$-In$_2$Se$_3$ was reported to behave as a metal.  It is also likely that the band gap of 1.38 eV determined from the onset in the photocurrent spectrum is larger than the calculated optical gap of 1.23 eV in the present work due to the high density of electrons in the conduction band in the experiment, leading to a blue shift due to the Moss$-$Burstein effect.

The VBM of the layered In$_2$Se$_3$ phases are all lower than that of CdSe. This can be explained by the $p$-$d$ coupling in CdSe which is likely to be stronger than in In$_2$Se$_3$. In CdSe, the filled Cd 4$d$ is well below the valence band composed of Se 5$p$ orbitals. The Cd 4$d$ $t_2$ states have the same symmetry as the VBM states, so the $p$-$d$ coupling pushes the VBM upwards.  Since In 4$d$ is lower than the Cd 4$d$, the $p$-$d$ coupling is expected to be weaker in In$_2$Se$_3$, resulting in lower VBM.

\section{Summary}

We reported on the electronic structure and optical properties of layered In$_2$Se$_3$ using the HSE06 hybrid functional with vdW corrections. We found that the fundamental band gaps are indirect, and the optical gaps are all larger than 1 eV.  In the case of $\beta$-In$_2$Se$_3$, which shares the same crystal structure of Bi$_2$Se$_3$, the fundamental band gap is only 0.49 eV, while the optical gap is 1.27 eV.  The small fundamental band gap of $\beta$-In$_2$Se$_3$ is not expected to lead to alloys of In$_2$Se$_3$ and Bi$_2$Se$_3$ with gaps much larger than the gaps of the parent compounds as suggested in the literature. The calculated absorption coefficient are found to exceed 10$^6$ cm$^{-1}$ and the onsets of optical absorption are overall in good agreement with experimental observations.

\section*{Acknowledgments}

We thank S. Law and Y. Wang for fruitful discussions. This work was supported by the National Science Foundation Faculty Early Career Development Program DMR-1652994. This research was also supported by the the Extreme Science and Engineering Discovery Environment supercomputer facility, National Science Foundation grant number ACI-1053575, and the Information Technologies (IT) resources at the University of Delaware, specifically the high performance computing resources.


\bibliography{2D-In2Se3}

\end{document}